\documentclass{PoS}

\usepackage{floatrow}

\title{Fermi acceleration under control: $\eta$ Carinae}

\ShortTitle{Fermi acceleration under control: $\eta$ Carinae}

\author{\speaker{Roland Walter}\\
        Department of Astronomy, University of Geneva, Switzerland\\
        E-mail: \email{roland.walter@unige.ch}}

\author{Matteo Balbo\\
        Department of Astronomy, University of Geneva, Switzerland\\
        E-mail: \email{matteo.balbo@unige.ch}}

\author{Christos Panagiotou\\
        Department of Astronomy, University of Geneva, Switzerland\\
        E-mail: \email{christos.panagiotou@unige.ch}}

\abstract{The $\eta$ Carinae binary system hosts one of the most massive stars, which features the highest known mass-loss rate. This dense wind encounters the much faster wind expelled by its stellar companion, dissipating mechanical energy in the shock, where particles can be accelerated up to relativistic energies and subsequently produce very-high-energy $\gamma$-rays. 
We used data from the Fermi Large Area Telescope obtained during the last 7 years and spanning two passages of $\eta$ Carinae at periastron and compared them with the predictions of particle acceleration in hydrodynamic simulations. Two emission components can be distinguished. The low-energy component cuts off below 10 GeV and its flux, modulated by the orbital motion, varies by a factor less than 2. Short-term variability occurs at periastron. The flux of the high energy component varies by a factor 3-4 but differently during the two periastrons. The variabilities observed at low-energy, including some details of them, and these observed at high-energy during the first half of the observations, do match the prediction of the simulation, assuming a surface magnetic field in the range 0.4-1 kG. The high-energy component and the thermal X-ray emission were weaker than expected around the second periastron suggesting a modification of the wind density in the inner wind collision zone. Diffuse shock acceleration in the complex geometry of the wind collision zone of $\eta$ Carinae provides a convincing match to the observations and new diagnostic tools to probe the geometry and energetics of the system. Orbital modulations of the high-energy component can be distinguished from these of photo absorption by the four large size telescopes of the Cherenkov Telescope Array to be placed in the southern hemisphere. e-Astrogam could easily discriminate between the lepto-hadronic and the hadronic models for the gamma-ray emission and constrain acceleration physics in more extreme conditions than found in SNR.}

\FullConference{35th International Cosmic Ray Conference --- ICRC2017\\
		10--20 July, 2017\\
		Bexco, Busan, Korea}

\begin{document}

\section{Introduction}

$\eta$ Carinae is the most luminous massive binary system of our galaxy and the first one to have been detected at very high energies, without hosting a compact object. It is composed by one among the most massive stars known ($\eta$ Car A) with an initial mass estimated above M$_{\rm{A}} \gtrsim 90 M_{\odot}$ \cite{2001ApJ...553..837H} and of a companion ($\eta$ Car B) believed to be an O supergiant or a WR star. 
$\eta$ Car A is accelerating a very dense wind with a mass loss rate of $\sim 8.5\cdot 10^{-4}$ M$_{\odot}$ yr$^{-1}$ and a terminal wind velocity of $\sim 420$ km s$^{-1}$ \cite{2012MNRAS.423.1623G}. Its companion probably emits a fast low-density wind at $10^{-5}$ M$_{\odot}$ yr$^{-1}$ reaching a velocity of 3000 km s$^{-1}$ \cite{2002A&A...383..636P,2005ApJ...624..973V,2009MNRAS.394.1758P}. 

During its Great Eruption (1837-1856), $\eta$ Carinae experienced a huge outburst ejecting an impressive quantity of mass estimated as $10-40~M_\odot$ \cite{2010MNRAS.401L..48G} at an average speed of $\sim 650$ km s$^{-1}$ \cite{2003AJ....125.1458S}, giving raise to the formation of the Homunculus Nebula, and became one of the brightest stars of the sky. The energy released in such a catastrophic event ($10^{49-50}$ erg) was comparable with a significant fraction of the energy emitted by a supernova explosion. The regular modulation detected in the X-ray lightcurves suggests that the two stars are located in a very eccentric orbit \cite{2001ApJ...547.1034C,2008MNRAS.388L..39O}. The estimated orbital period at the epoch of the Great Eruption was $\sim$~5.1 yr, and increased up to the current $\sim~5.54$~yr \cite{2004MNRAS.352..447W,2005AJ....129.2018C,2008MNRAS.384.1649D}. 

Given the high eccentricity of the orbit, the relative separation of the two stars varies by a factor $\sim20$, reaching its minimum at periastron, when the two objects pass within a few AU of each other (the radius of the primary star is estimated as 0.5 AU). In these extreme conditions their supersonic winds interact forming a colliding wind region of hot shocked gas where charged particles can be accelerated via diffusive shock acceleration up to high energies \cite{1993ApJ...402..271E,2003A&A...409..217D,2006ApJ...644.1118R}. As these particles encounter conditions that vary with the orbital phase of the binary system, one can expect a similar dependency in the $\gamma$-ray emission.

The hard X-ray emission detected by INTEGRAL \cite{2008A&A...477L..29L} and Suzaku \cite{2008MNRAS.388L..39O}, with an average luminosity $(4$-$7)\times10^{33}$ erg s$^{-1}$, suggested the presence of relativistic particles in the system. The following year AGILE detected a variable source compatible with the position of $\eta$ Carinae \cite{2009ApJ...698L.142T}. Other $\gamma$-ray analyses followed, reporting a luminosity of $1.6\times10^{35}$ erg s$^{-1}$ \cite{2010ApJ...723..649A,2011A&A...526A..57F,2012A&A...544A..98R}, and suggested the presence of a hard component in the spectrum around periastron, which subsequently disappeared around apastron. Such a component has been explained through $\pi^0$-decay of accelerated hadrons interacting with the dense stellar wind \cite{2011A&A...526A..57F}, or interpreted as a consequence of $\gamma$-ray absorption against an ad hoc distribution of soft X-ray photons \cite{2012A&A...544A..98R}.

Cherenkov observations \cite{2012MNRAS.424..128H} did not yet led to any significant detection, providing an upper limit at energies $\gtrsim 500$ GeV that implies a sudden drop in the $\gamma$-ray flux, which could be related to a cut-off in the accelerated particle distribution or to severe $\gamma-\gamma$ absorption.

\section{Variability in the GeV band}

We have analysed the Fermi LAT data in two energy bands (0.3-10 GeV and 10-300 GeV). The details of the analysis can be found in \cite{2017arXiv170502706B}. The position of the $\gamma$-ray source matches perfectly the nominal coordinates of $\eta$ Carinae, considering the complete 7-year data set or shorter time intervals.

\begin{figure}[]
\begin{minipage}{\textwidth}
\floatbox[{\capbeside\thisfloatsetup{capbesideposition={right,top},capbesidewidth=0.5\textwidth}}]{figure}[\FBwidth]
{\caption{Simulated and observed X-ray and $\gamma$-ray lightcurves of $\eta$ Carinae. The black and purple lines and bins show the predicted inverse-Compton and neutral pion decay lightcurves. The green and red points show the observed Fermi-LAT lightcurves at low (0.3-10 GeV) and high (10-300 GeV) energies. The dim grey lightcurves show the observed (continuous) and predicted (dash, without obscuration) thermal X-ray lightcurves. Error bars are $1\sigma$.}\label{fig:simul}}
{\includegraphics[width=0.5\textwidth]{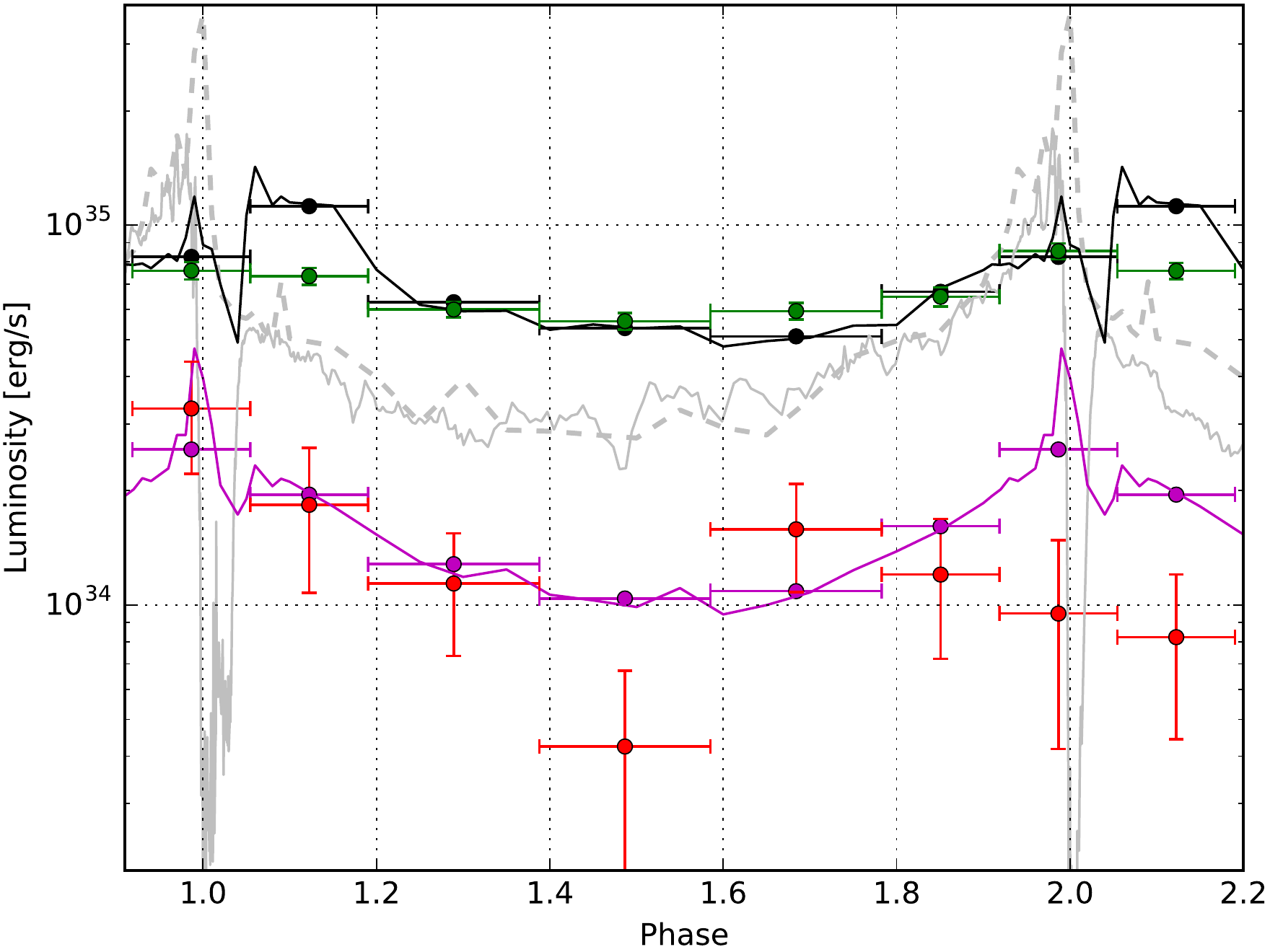}}
\end{minipage}
\end{figure}

The lightcurve of the high-energy flux of $\eta$ Carinae obtained from the likelihood analysis is reported in Fig.~\ref{fig:simul}. After the first periastron passage of 2009 the flux of $\eta$ Carinae decreased slightly towards apastron. The flux did however not increase again toward the periastron of 2014. 

\begin{figure}[]
\begin{minipage}{\textwidth}
\floatbox[{\capbeside\thisfloatsetup{capbesideposition={right,top},capbesidewidth=0.5\textwidth}}]{figure}[\FBwidth]
{\caption{A merged Fermi LAT analysis (0.3-10 GeV) of the two periastrons for narrow time bins. The two broad bins and the black curve are the same as in Fig. \ref{fig:simul}.}\label{fig:peri}}
{\includegraphics[width=0.5\textwidth]{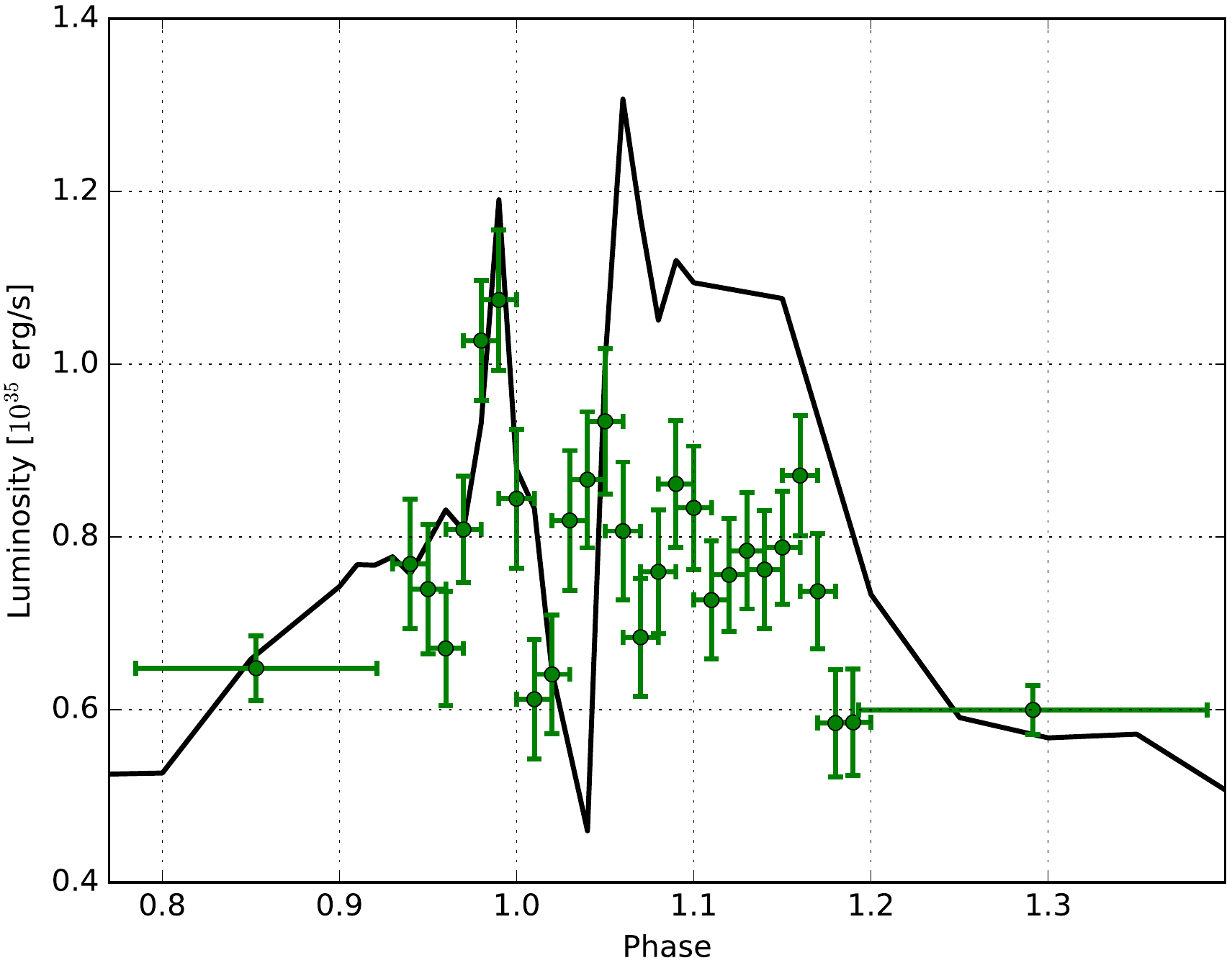}}
\end{minipage}
\end{figure}

In the low-energy band, we performed the analysis assuming a power law spectrum with an exponential cutoff, which represent well the averaged spectrum. The low-energy $\gamma$-ray variability appears similar for the two periastrons (see Fig.~\ref{fig:simul}), and, as the spectra are compatible, we used the good statistics available to perform a merged analysis of these two periods on shorter time bins (Fig.~\ref{fig:peri}).

\cite{2011ApJ...726..105P} presented three dimensional hydrodynamical simulations of $\eta$ Carinae including radiative driving of the stellar winds \cite{1975ApJ...195..157C}, optically-thin radiative cooling \cite{2000adnx.conf..161K}, gravity and orbital motion. The main aim of these simulations was to reproduce the X-ray emission analysing the emissivity and the self-obscuration of the stellar wind. The simulations reproduced the observed X-ray spectra and lightcurves reasonably well, excepting the post-periastron extended X-ray minimum, where flux was overestimated. Additional gas cooling, e.g by particle acceleration and inverse-Compton processes, could decrease the wind speed and increase the cooling and disruption of the central wind collision zone.

To estimate the non thermal emission we first calculated the maximum energies that could be reached by electrons and hadrons \cite{2011A&A...526A..57F} cell-by-cell assuming a dipolar magnetic field at the surface of the main star, perpendicular to the orbital plane (reality is more complex with the two stars contributing). The magnetic field is the only additional parameter and can be tuned. We calculated shock velocities and mechanical power in every cell, including those one outside the shock region. As expected, most of the shock power is released on both sides of the wind collision zone and in the cells downstream the wind-collision region \cite{2006ApJ...644.1118R}. The increasing shock area compensates the loss of the released energy density up to a relatively large distance from the center of mass, explaining why the X-ray luminosity at apastron is about a third of the peak emission at periastron. 

The energy available in electrons and hadrons were then summed in the ranges 0.3 $< \rm{E}_e <$ 10 GeV and $\rm{E}_p>20$ GeV, respectively to match the spectral bands observed by Fermi-LAT. The local cell physical properties can be used to easily estimate pion production as long as the Larmor radius is similar to the cell size. The minimum size of the cells in the simulation, $\sim 10^{11}$ cm, is larger than the proton Larmor radius for Lorentz gamma factor up to 10$^5$. Only one third of the power accelerating protons is available to produce $\gamma$-rays through the neutral pion channel. Electron cooling and pion decay occur instantaneously when compared to other time scales. 

To consider the possible effects of photon-photon opacity we calculated the X-ray thermal emission in each cell, and evaluated the optical depth along different lines of sight. As the current orientation of the binary system, with respect to the Earth, still presents quite some uncertainties \cite{2012ApJ...746L..18M}, we used several of them which provided optical depth $\tau$ varying between 10$^{-6}$ at apastron and $\sim$10$^{-2}$ at periastron. This excludes explaining the 1-100 GeV spectral shape by the effects of photon-photon absorption \cite{2012A&A...544A..98R}.

The mechanical luminosity available in the shock increases towards periastron (the same trend is followed by the thermal emission) and almost doubles in the phase range $\approx 1.05 - 1.15$. The latter peak corresponds to a bubble with reverse wind conditions developing because of the orbital motion, effectively doubling the shock front area during about a tenth of the orbit \cite{2011ApJ...726..105P}. The density of this bubble is low so its thermal emission does not contribute significantly to the X-ray lightcurve. The mechanical luminosity shows a local minimum between phases 1.0 and 1.05, when the central part of the wind collision zone is disrupted. 

Electron cooling, through inverse-Compton scattering, is very efficient and such $\gamma$-rays are expected to peak just before periastron. A secondary inverse-Compton peak could be expected above phase 1.05 although its spectral shape could be very different as the UV seed thermal photons will have lower density when compared to the location of the primary shock close to the center of the system. In our simplified model we assumed that the spectral shape of the seed photons is the same in all cells of the simulation (r$^{-2}$ dependency is taken into account), and that these soft photons are enough to cool down all the relativistic electrons. The relative importance of the second peak, however, depends on the magnetic field geometry, radiation transfer (neglected in our model), obscuration and details of the hydrodynamics (which do not represent the soft X-ray observations very well in this phase range). These details are not well constrained by the available observations and we did not try to refine them. 

The situation is different for hadrons. Unless the magnetic field would be very strong ($>$ kG) hadronic interactions mostly take place close to the center and a single peak of neutral pion decay is expected before periastron.

Figure \ref{fig:simul} shows the X and $\gamma$-ray lightcurves predicted by the simulations for a magnetic field of 500 G and assuming that 1.5\% and 2.4\% of the mechanical energy is used to respectively accelerate electrons and protons. To ease the comparison between observations and simulations, the results of the latter were binned in the same way as the observed data.

Both the predicted inverse-Compton emission and the observed (0.3-10 GeV) LAT lightcurve show a broad peak extending on both sides of periastron, as expected from the evolving shock geometry. The amplitude of the variability in the simulation depend on the number/size of those cells where particles can be accelerated up to relevant energies, which in turn depends on the magnetic field. Probing the range suggested by \cite{2012SSRv..166..145W}, a surface magnetic field larger than 400~G provides a good match to the observations, while lower fields produce too large variations. In this work we have not considered any magnetic field amplification at the shock, which in turn could obviously scale down the surface magnetic field required to get equivalent results. Assuming a field of 500 G for the rest of the discussion, the predicted flux at phase 1.1 is twice too large when compared with the observation. This discrepancy largely comes from the energy released in the inverted wind bubble after periastron. The ratio of the emission generated in the shocks on both sides of the wind collision zone is relatively constant along the orbit excepting at phase 1.1, where much more power is generated in the shock occurring in the wind of the secondary star. The inverted bubble might either be unstable in reality or produce a significantly different inverse-Compton spectrum.

Relativistic electrons immersed in such a high magnetic field will produce a synchrotron radiation at low energy. Knowing the inverse-Compton spectrum, we can estimate the synchrotron peak luminosity. Around apastron it is several order of magnitude ($\sim$10$^6$) fainter than the inverse-Compton peak. Close to periastron the synchrotron emission peaks in the optical band, at a level two orders of magnitude fainter than the inverse-Compton peak. Those limits are in agreement with the estimated radio upper limit \cite{2003MNRAS.338..425D}.

Since the low energy spectra during both periastrons are sufficiently in agreement, we analysed simultaneously the Fermi LAT low energy data derived from the two periastrons, binned in shorter time intervals (Fig.~\ref{fig:peri}). They show a peak at periastron, a minimum at phase 1.02 and a second broad peak at phase 1.1. It is very similar to the prediction of the simulation for the inverse-Compton luminosity. The only notable exception is that the observed second broad peak is slightly shifted towards earlier phases and has a lower luminosity when compared to the simulation. The similarities between the observations and the simulation, $\gamma$-ray peak and minimum with consistent duration and amplitude, are very encouraging. The phase difference could be related to the eccentricity $(\epsilon=0.9)$ assumed in the simulation, which is not well constrained observationally \cite{2000ApJ...528L.101D,2001ApJ...547.1034C} and that has an important effect on the inner shock geometry. 

The distribution of $\gamma_e$, weighted by the emissivity, is relatively smooth and the expected photon distribution is very smooth. The difference of the electron spectral shape on both sides of the wind collision zone cannot explain the two components $\gamma$-ray emission as suggested by \cite{2011A&A...530A..49B}, who assumed a simplified geometry. We obtain a good match between the observed low energy $\gamma$-ray spectrum and the predictions of the simulations at periastron, even though some discrepancy can be observed at apastron where an excess is observed between 2 and 10 GeV.

The inverse-Compton emission peaks slightly below 1 GeV and does not extend beyond 10 GeV at a level consistent with the observations during the first periastron, contrasting with the conclusions from \cite{2015MNRAS.449L.132O}, attributing the full Fermi LAT detection to hadronic emission. Their simulations predict a smaller variation between periastron and apastron, a longer flare around periastron and a deeper minimum when compared to the observed data. Such discrepancies might be due to the simplified geometry assumed by the authors and by the artificially reduced particle acceleration at periastron. Inverse-Compton emission and neutral pion decay \cite{2011A&A...526A..57F} remains therefore a very good candidate to explain the Fermi observations.

The simulated pion induced $\gamma$-ray lightcurve and its variability amplitude show a single peak of emission centered at periastron, in good agreement with the Fermi LAT observations of the first periastron. The results of the observations of the second periastron are different, with a lack of emission. It has been suggested that the change of the X-ray emission after that periastron (a significant decrease can be observed in Fig. \ref{fig:simul}, see also  \cite{2015arXiv150707961C}) was the signature of a change of the wind geometry, possibly because of cooling instabilities. A stronger disruption or clumpier wind after the second periastron could perhaps induce a decrease of the average wind density and explain that less hadronic interactions and less thermal emission took place, without affecting  much inverse-Compton emission.

Proton could be accelerated up to $10^{15}$ eV around periastron and reach $10^{14}$ eV on average. 
The choice of a lower magnetic field reduce those energies at apastron to $\sim 6 \times 10^{12}$~eV and $\sim 2 \times 10^{12}$~eV, and at periastron to $\sim 5.6 \times 10^{14}$~eV and $\sim 1.9 \times 10^{14}$~eV, respectively for 300~G and 100~G. $\eta$ Carinae can therefore probably accelerate particles close to the knee of the cosmic-ray spectrum. The spectra and the maximum particle energy depend of course on several assumptions, in particular the magnetic field. The highest energy $\gamma$-rays will be photo-absorbed and orbital modulation could be expected in the TeV domain. The duration of the periastron bin [0.92-1.06] corresponds to more than 260 days and is longer than the interaction timescale of the protons responsible for the flux variability.

Assuming that for each photon originated via hadronic processes we also have the production of one neutrino, we derive a neutrino flux above 10 TeV that might reach $10^{-9}$ GeV s$^{-1}$cm$^{-2}$ on average, which is of the order of the IceCube neutrino sensitivity for several years of observations \cite{2017ApJ...835..151A}. Stacking some months of IceCube data obtained around periastron, over several decades could allow the detection of one PeV neutrino, above the atmospheric background. 

A strong $\gamma$-ray variability is expected above 100 GeV. Depending on the assumed soft energy photons distribution and the consequent $\gamma$-$\gamma$ absorption at very high energy, $\eta$ Carinae could be detected by the CTA southern array (including four large size telescopes) at more than $10\sigma$ in spectral bins of $\Delta {\rm E/E} = 20\%$ for exposures of 50 hours, enough to measure separately the variability along the orbit of the high energy component and of photo absorption \cite{Acharya20133}. 

$\gamma$-ray observations can probe the magnetic field and shock acceleration in details, however the quality of the current data above 1 GeV does not yet provide enough information to test hydrodynamical models including detailed radiation transfer (inverse-Compton, pion emission, photo-absorption). The interplay between disruption and obscuration does not yet account for the X-ray minimum and orbit to orbit variability. More sensitive $\gamma$-ray observations will provide a wealth of information and allow to test the conditions and the physics of the shocks at a high level of details, making of $\eta$ Carinae a perfect laboratory to study particle acceleration in wind collisions. $\eta$ Carinae could yield to $10^{48-49}$ erg of cosmic-ray acceleration, a number close to the expectation for an average supernova remnant \cite{2016APh....81....1B}.

\section{Spectral energy distribution and acceleration physics}

The spectral energy distribution of $\eta$ Carinae features an excess of emission at hard X-rays, beyond the extrapolation of the thermal emission \cite{Panagiotou2017} that should connect to the Fermi spectrum in a yet unknown manner.

\begin{figure}[]
\begin{minipage}{\textwidth}
\floatbox[{\capbeside\thisfloatsetup{capbesideposition={right,top},capbesidewidth=0.3\textwidth}}]{figure}[\FBwidth]
{\caption{Spectral energy distribution of $\eta$ Carinae from 1 keV to 10 GeV. The data are from NuStar (grey), Swift/BAT (cyan), INTEGRAL (purple), Fermi/LAT (black) and the upper limits from HESS (green). The predictions are from mostly hadronic (dashed blue line) \cite{2015MNRAS.449L.132O} and lepto-hadronic (red line for the leptonic part) \cite{2017arXiv170502706B} models. The sensitivity curves of e-Astrogam (in the galactic plane) and CTA are also indicated (dotted yellow lines).}\label{fig:sed}}
{\includegraphics[width=0.65\textwidth]{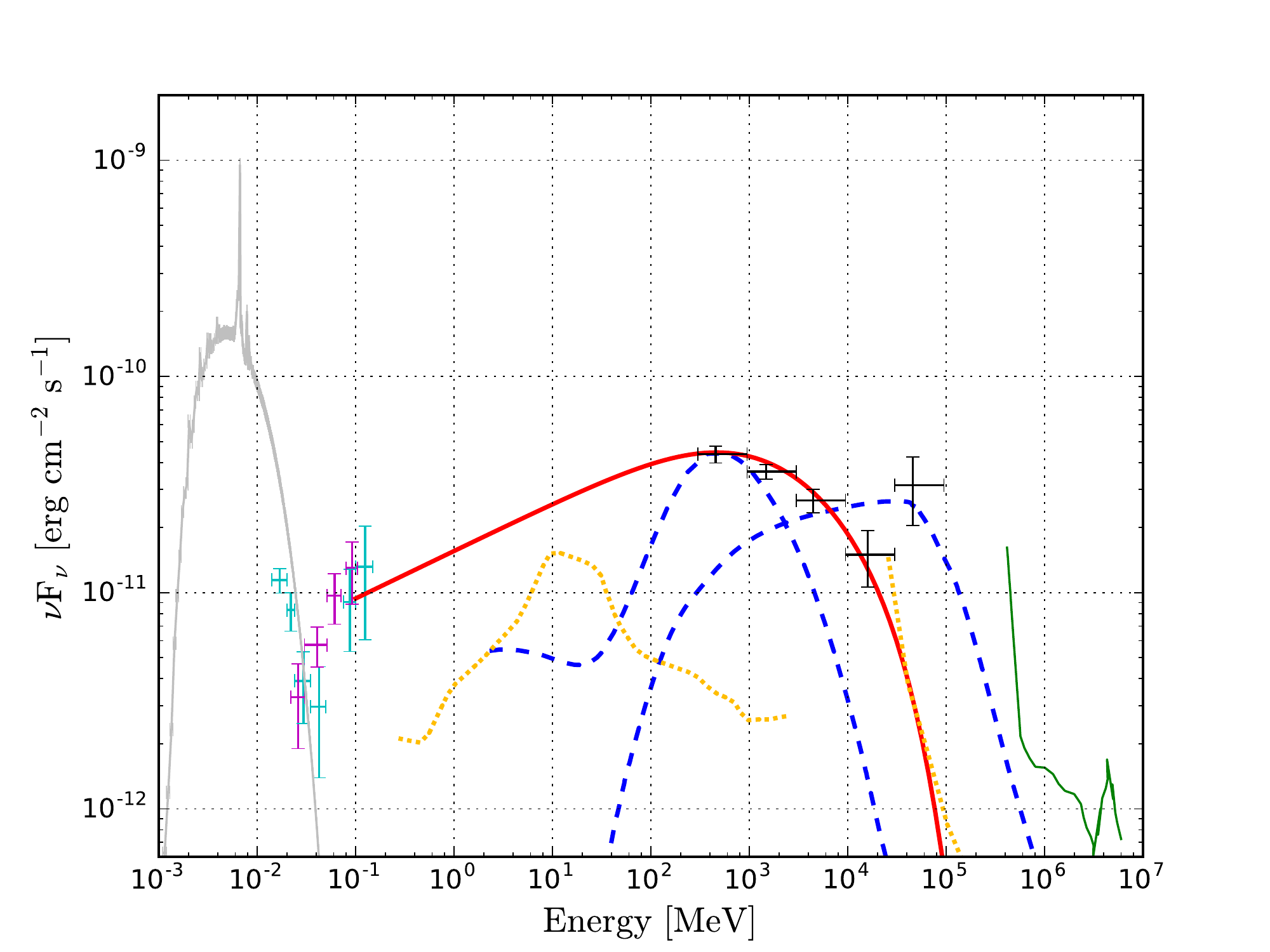}}
\end{minipage}
\end{figure}

In the above section we have presented a model where electrons and protons are accelerated (as initially proposed by \cite{1993ApJ...402..271E}). The fraction of the shock mechanical luminosity accelerating electrons appears to be slightly smaller than the one that accelerates protons. 
These results contrast with the efficiencies derived from the latest particle-in-cell 
simulations \cite{2015PhRvL.114h5003P}, involving low magnetic fields, radiation and particle densities and favouring hadronic acceleration in the context of SNR. 

Purely hadronic acceleration has been proposed \cite{2015MNRAS.449L.132O} to explain the GeV spectrum of Eta Carinae. In that case the two spectral components are related to the different hadron interaction times observed on the two sides of the wind separation surface, largely because of the contrast in density and magnetic field. In our simulations this effect is smoothed by the many zones of the model, each characterized by different conditions. Even if the shock on the companion side does contribute more at high energies, the resulting pion decay spectrum does not feature two components.

GeV-keV photoabsorption can in principle affect the observed spectral shape and create a local minimum close to 10 GeV \cite{2012A&A...544A..98R}. As the relevant optical depth remains negligeable, this is however unlikely to play a significant role \cite{2017arXiv170502706B} and would, in addition, require an excessively large cutoff energy.

An instrument sensitive in the 1-100 MeV band, such as e-Astrogam \cite{2017ExA...tmp...24D}, will easily discriminate between the lepto-hadronic and the hadronic models for the gamma-ray emission as the inverse Compton leptonic emission of the former would be much stronger than predicted by the latter (Fig. \ref{fig:sed}). e-Astrogam can therefore decide which is the model likely to explain the high energy emission of Eta Carinae and strongly constrain the acceleration physics in more extreme conditions than found in SNR.

\bibliography{references}

\end{document}